\documentclass[12pt]{iopart}

\usepackage{iopams}
\usepackage{graphicx}
\usepackage{dcolumn}
\usepackage{bm}

\begin{document}

\title[Stationary waves in a superfluid exciton gas in quantum Hall bilayers]
{Stationary waves in a superfluid exciton gas in quantum Hall
bilayers}

\author{A. A. Pikalov, D. V. Fil}

\address{Institute for Single Crystals National Academy of
Sciences, Lenin ave. 60, Kharkov 61001, Ukraine}
\ead{fil@isc.kharkov.ua}
\begin{abstract}
Stationary waves in a superfluid magnetoexciton gas in $\nu=1$
quantum Hall bilayers are considered. The waves are induced by
counter-propagating electrical currents that flow in a system with
a point obstacle. It is shown that stationary waves can emerge
only in imbalanced bilayers in a certain diapason of currents. It
is found that the stationary wave pattern is modified
qualitatively under a variation of the ratio of the interlayer
distance to the magnetic length  $d/\ell$. The advantages of use
graphene-dielectric-graphene sandwiches for the observation of
stationary waves are discussed. We determine the range of
parameters (the dielectric constant of the layer that separates
two graphene layers and the ratio $d/\ell$) for which the state
with superfluid magnetoexcitons can be realized in such
sandwiches. Typical stationary wave patterns are presented as
density plots.
\end{abstract}


\vspace{2pc}

\submitto{\JPCM}

\section{Introduction}

Realization of superfluidity of excitons in semiconductor
heterostructures is considered as a challenge problem. Exciton
mass is smaller or of order of the electron mass and the
temperature of the transition into a superfluid state can be quite
large. Considerable attention has been given to the superfluidity
of spatially indirect excitons in bilayer electron structures. A
flow of such excitons results in an appearance of counterflow
electrical currents in the layers. They can provide feeding of the
load located at one end of the system from the source located at
the other end, and superfluidity of spatially indirect excitons
can be considered as a kind of unconventional superconductivity.
The idea goes back to the papers \cite{ll,sh}. This idea was
reincarnated in the quantum Hall physics of bilayer electron
systems in Refs. \cite{h1,h2,h3}. In a quantum Hall bilayer with
the total filling factor of Landau levels $\nu_T=1$ the state with
a spontaneous interlayer phase coherence may emerge (see
\cite{1,2,3,bez,n4} as a review). The many-particle wave function
for such a state describes the pairing of electrons from one layer
and holes from the other layer and its structure is analogues to
the BCS wave function. Experimental studies of transport
properties of quantum Hall bilayers \cite{4,5,6,7,8} confirm in
some part the theoretical prediction. The systems demonstrate
vanishing of the Hall resistance and exponential increase of the
counterflow conductivity under lowering of temperature. But the
conductivity remains finite that can be accounted for unbound
vortices \cite{9,10,11}.

Another hallmark of superfluidity is that an obstacle does not
disturb a flowing superfluid at small flow velocity. If the flow
velocity exceeds the critical one, the obstacle produces excitations
that are in rest relative to it. They are the stationary waves - the
waves with the time independent phase in each space point (in the
reference frame connected with the obstacle). Stationary waves may
emerge in a medium which excitation spectrum has a dispersion. The
well-known example is ship waves \cite{12} - the waves produces by a
ship on the water surface. The spectrum of deep water waves has
negative dispersion and ship waves on a deep water are located
behind the floating ship. The Bogolyubov spectrum has positive
dispersion and stationary waves in a superfluid weakly non-ideal
Bose gas are located outside the Mach cone: the waves appear ahead
the obstacle, and do not appear strictly behind it \cite{13}.

Stationary waves have been recently observed  in an
exciton-polariton gas \cite{14}. Since the exciton-polariton mass
is much
 smaller than the electron mass,
 a two-dimensional exciton-polariton gas in microcavities is
considered as a candidate for a high-temperature superfluid
\cite{15, 16,17}. The observation \cite{14} consists in that at
rather high exciton-polariton concentration and small flow
velocity the waves are not excited.  But if the flow velocity
exceeds the speed of sound  stationary waves emerge outside the
Mach cone.

In this paper we consider the possibility of observation of
stationary waves  in quantum Hall bilayers in a state with a
spontaneous interlayer phase coherence. In Sec. \ref{s2} we use
the kinematic approach \cite{12} based on the analysis of the
excitation spectrum and determine the range of the filling factor
imbalance and the counterflow currents where stationary waves may
emerge. It is found that the stationary wave pattern is modified
qualitatively under variation of the ratio of the interlayer
distance $d$ to the magnetic length $\ell$. In Sec. \ref{s3} we
present the description of stationary waves in $\nu_T=1$ quantum
Hall bilayers in  the dynamical approach \cite{12}. The approach
allows to compute the amplitudes and phases of stationary wave in
each space point. The results are presented as density plots. In
Sec. \ref{s4} we discuss advantages of use graphene bilayers for
the observation of stationary waves.

\section{The excitation spectrum and conditions for the emergence of
stationary waves in quantum Hall bilayers} \label{s2}

In this section we consider under what conditions  stationary
waves can be excited in a superfluid magnetoexciton gas in
bilayers.

We use the kinematic approach \cite{12} that is based on the
analysis of the excitation spectrum.  The spectrum at general
direction of the wave vector relative to the direction of
counterflow currents and general filling factor imbalance was
obtained in \cite{18,19} basing on the approach \cite{20}. Here we
reproduce in short the derivation \cite{19}.

The state with the spontaneous interlayer phase coherence is
described by the many-particle wave function \cite{1}
\begin{equation}  \label{dcs}
|\Psi\rangle=\prod_X\left(\cos\frac{\theta_X}{2}a^+_{1,X}+e^{iQ_x
 X+
 i\tilde{\varphi}_X}\sin\frac{\theta_X}{2}a^+_{2,X-Q_y\ell^2}\right)|0\rangle.
\end{equation}
Here we use the form of (\ref{dcs}) given in Ref. \cite{19}. It
corresponds to a coordinate system with the $x$ axis directed at
some angle to the counterflow current direction. The quantities
$a^+_{i, X}$, $a_{i,X}$ are the creation and annihilation
operators for the electrons in zeroth Landau level (the lowest
Landau level approximation is used), $i$ is the layer index, and
$X$ is the guiding center coordinate. The vector-potential is
chosen in the form ${\bf A}=(0, B x,0)$ that corresponds to the
single particle wave function $\psi_X({\bf
r})=(1/\pi^{1/4}\ell)e^{i X y/ \ell^2}\exp[-(x-X)^2/2\ell^2]$. The
variable $\theta_X$ can be presented as a sum of the mean value
and the fluctuating part: $\theta_X=\theta_0+\tilde{\theta}_X$.
The mean value $\theta_0$ is connected with the filling factors by
the relation $\cos \theta_0=\nu_1-\nu_2$, where  $\nu_i=2\pi n_i
\ell^2$ is the filling factor for the $i$-th layer, and $n_i$ is
the electron concentration. The phase contains the regular $Q_x X$
and the fluctuating $\tilde{\varphi}_X$ parts.

 The
counterflow supercurrents are directed along the vector ${\bf
Q}=(Q_x,Q_y)$. Their values are given by the equation
\begin{equation}\label{jj}
    {\bf j}_{CF}=\frac{e}{\hbar}\frac{1}{S}\frac{d E_{0}}{d
    {\bf Q}}.
\end{equation}
The quantity ${\bf j}_{CF}$ is defined as the density of the
supercurrent in the layer 1 (${\bf j}_{1}=-{\bf j}_{2}={\bf
j}_{CF}$), $E_{0}=\langle \Psi_0 |H_C |\Psi_0\rangle$ is the
mean-field energy for  the Hamiltonian of Coulomb interaction in
zeroth Landau level in the state (\ref{dcs}) at
$\tilde{\theta}_X=\tilde{\varphi}_X=0$, and $S$ is the area of the
system.

The explicit expression for $E_{0}$ reads as
\begin{equation}\label{e-mf}
    E_{0}=\frac{S}{8\pi \ell^2}\left[\cos^2 \theta_0
    [H_0-F_S(0)]
    -\sin^2 \theta_0 { F}_D(Q)\right],
\end{equation}
where the quantities
\begin{eqnarray} \label{fd}
F_{S(D)}(q)=\frac{1}{2\pi}\int_0^{\infty}p J_0(p q \ell^2)
V_{S(D)}(p)e^{-\frac{p^2 \ell^2}{2}}d p
\end{eqnarray}
($J_0(x)$ is the Bessel function) describe the contribution of the
intralayer (S) and the interlayer (D) exchange  Coulomb
interaction, and the quantity
\begin{equation}\label{h0}
    H_0=\lim_{q\to 0}\frac{V_S(q)-V_D(q)}{2\pi\ell^2},
\end{equation}
 the contribution of the direct  Coulomb interaction.
  $V_S(q)=2\pi e^2/\varepsilon q$
 and $V_D(q)=2\pi e^2 e^{- q d}/\varepsilon q$ are the Fourier
 components of the interaction, and $\varepsilon$ is the
 dielectric constant.

In the harmonic approximation the energy of  fluctuations  has the
form
\begin{eqnarray} \label{9}
E_{2}=\sum_q [m_z(-q)K_{zz}(q)m_z(q)+
\frac{1}{4}\varphi(-q)K_{\varphi\varphi}(q)\varphi(q)
\cr-\frac{1}{2}(im_z(-q)K_{z\varphi}(q)\varphi(q)+c.c.)],
\end{eqnarray}
where
\begin{eqnarray} \label{a7}
{m} _ {z} (q) = \frac{1}{2}\sqrt{\frac {2 \pi l^2} {S}} \sum_X
\left(\cos\theta_X-\cos \theta_0\right) e ^ {-{i} q X}, \cr
{\varphi} (q) = \sqrt{\frac {2 \pi l^2} {S}} \sum_X
\tilde{\varphi} (X) e ^ {-{i} q X}.
\end{eqnarray}
The components of the matrix $\bm{K}$ are defined as
$$K_{\alpha\beta}(q)=K_{\alpha\beta}({\bf q})\Big|_{{\bf q}=q \mathbf{i}_x},$$
where
\begin{eqnarray} \label{Ks}
K_{zz}({\bf q})=H({\bf q},{\bf Q})-F_S(|{\bf q}|)+F_D(|{\bf
Q}|)\cr+\left(F_D(|{\bf Q}|) - \frac{F_D(|\mathbf{q}
+\mathbf{Q}|)+F_D(|\mathbf{q}
-\mathbf{Q}|)}{2}\right)\cot^2\theta_0,\cr K_{\varphi\varphi}({\bf
q})=\sin^2\theta_0 \left[F_D(|{\bf Q}|)- \frac{F_D(|\mathbf{q}
+\mathbf{Q}|)+F_D(|\mathbf{q} -\mathbf{Q}|)}{2}\right],\cr
K_{z\varphi}({\bf q})=-\cos\theta_0 \frac{F_D(|\mathbf{q}
+\mathbf{Q}|)-F_D(|\mathbf{q} -\mathbf{Q}|)}{2},
\end{eqnarray}
and
\begin{eqnarray} \label{HFs}
H({\bf q},{\bf Q})=\frac{1}{2\pi\ell^2}\left[V_S(|{\bf
q}|)-V_D(|{\bf q}|)\cos\left(|\mathbf{q}\times\mathbf{Q}|
\ell^2\right)\right]e^{-\frac{q^2 \ell^2}{2}}.
\end{eqnarray}

The quantities $\hbar m_z(q)$  and $\varphi(-q)$ are the
conjugated variables. It can be checked by computing the
commutator of the operators that correspond to these variables.
The canonical equations of motion read as
\begin{eqnarray} \label{13}
\hbar\frac{d {\varphi}(q)}{d t}= 2 K_{zz} (q) {m}_z(q)- i
K_{z\varphi}(q) {\varphi}(q) ,\cr \hbar\frac{d {m}_z(q)}{d t}=-
\frac{1}{2}K_{\varphi\varphi}(q) {\varphi}(q)  -  i
K_{z\varphi}(q) {m}_z(q).
\end{eqnarray}
According to Eq. (\ref{13}) the energy of the collective mode with
the wave vector ${\bf q}=q {\bf i}_x$ has the form
$\Omega(q)=\sqrt{K_{\varphi\varphi}(q)K_{zz}(q)}+K_{z\varphi}(q)$.
Rotating the axes we find the excitation spectrum at general ${\bf
q}$
\begin{equation}\label{e1}
    \Omega({\bf q})=\sqrt{K_{\varphi\varphi}({\bf q})K_{zz}({\bf q})}
    +K_{z\varphi}({\bf q}).
\end{equation}
In what follows we use the reference frame with the $x$ axis
directed along ${\bf Q}$ (${\bf Q}=(Q,0)$) that coincides with the
direction of the current ${\bf j}_{CF}$ Eq. (\ref{jj}).

The wave vectors for the stationary waves satisfy the equation
\begin{equation}\label{st1}
     \Omega({\bf q}_{st})=0.
\end{equation}
The necessary condition for the observation of stationary waves is
the existence of nontrivial solutions of Eq. (\ref{st1}). Beside
that, the spectrum $\Omega({\bf q})$ should be real valued at all
${\bf q}$ (complex valued $\Omega({\bf q})$ signal for the
dynamical instability of the state Eq. (\ref{dcs})). These two
conditions determine the parameters of the systems where
stationary waves can emerge.

\begin{figure}
  \begin{center}
   \includegraphics{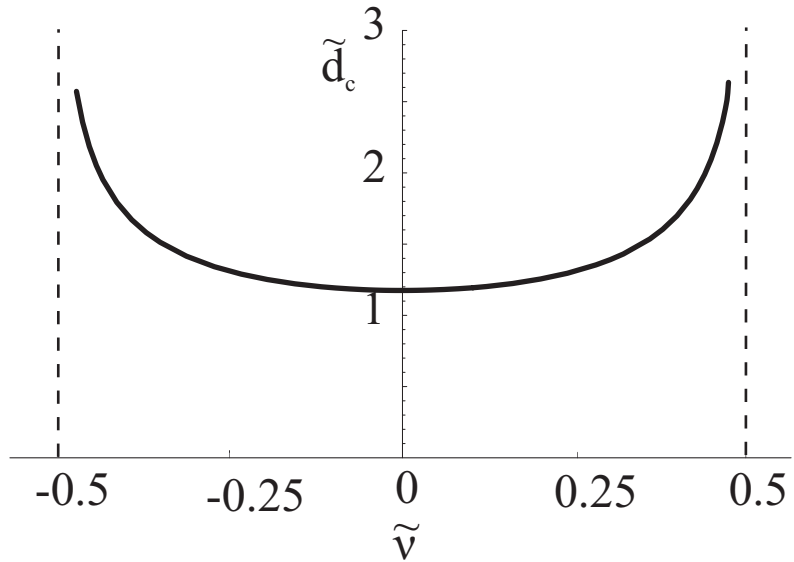}\\
  \caption{Critical interlayer distance versus
  the filling factor imbalance}\label{fig1}
  \end{center}
\end{figure}

\begin{figure}
\begin{center}
 \includegraphics[width=10
 cm]{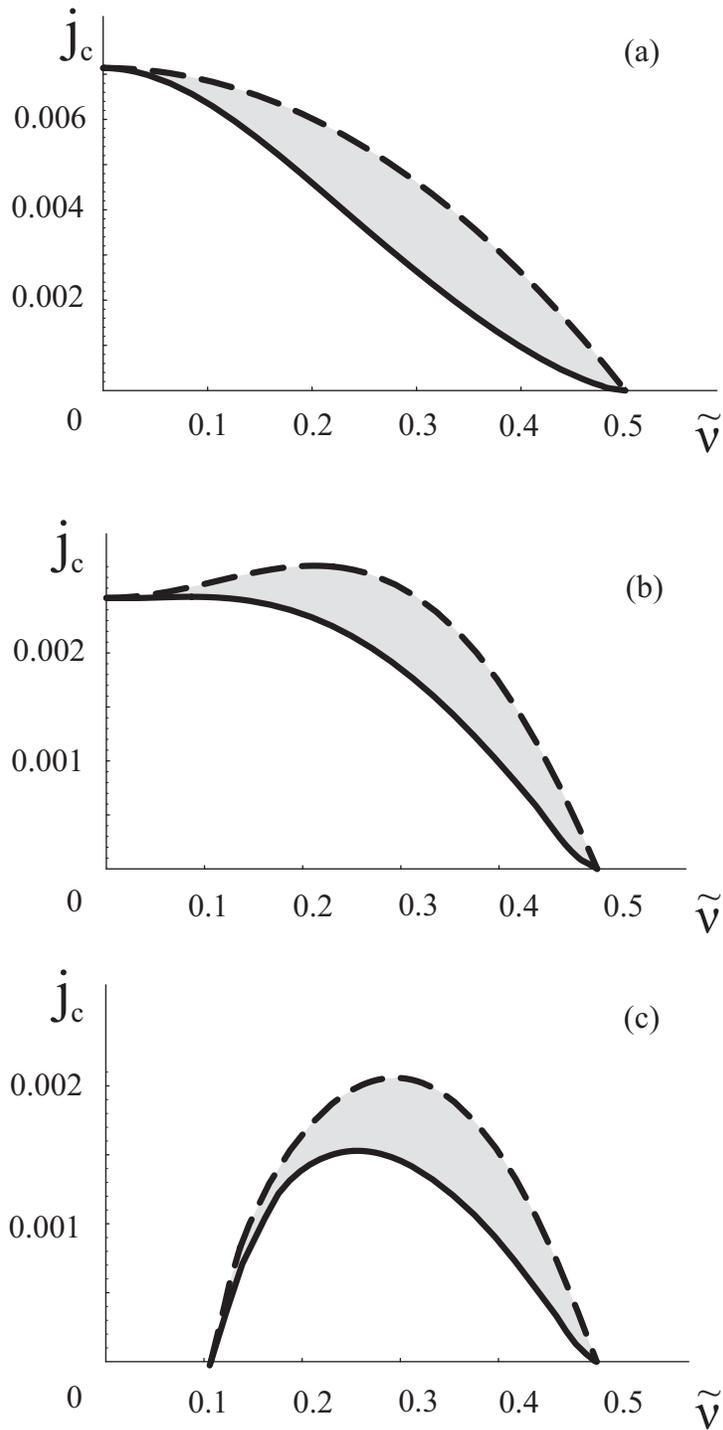}\\
  \caption{The critical currents $j_{c1}$
  (solid line) and
  $j_{c2}$ (dashed line) at $d/\ell=0.3$ (a), $d/\ell=1.1$ (b), and $d/\ell=1.2$ (c).
  The current density $j$ is in $e^3/\hbar\varepsilon \ell^2$ units. The
  range of currents and filling factor imbalance
  where
  stationary waves are excited is shown grey.
   }
 \label{fig2}
\end{center}
\end{figure}

The spectrum (\ref{e1}) depends on three parameters: the ratio
$\tilde{d}=d/\ell$, the filling factor imbalance
$\tilde{\nu}=(\nu_1-\nu_2)/2$ and the gradient of the phase $Q$.
The spectrum should be real valued at $Q=0$. It gives the
restriction $\tilde{d}<\tilde{d}_c(\tilde{\nu})$. The dependence
$\tilde{d}_c(\tilde{\nu})$ is shown in Fig. \ref{fig1}. The
stationary wave equation (\ref{st1}) has nontrivial solutions at
$Q>Q_1(\tilde{d},\tilde{\nu})$. On the other hand, the condition
for the quantity $\Omega({\bf q})$ be real valued at all ${\bf q}$
yields the restriction $Q<Q_2(\tilde{d},\tilde{\nu})$ ($Q_1\leq
Q_2$) Thus, stationary waves can be observed at $Q_{c1}<Q<Q_{c2}$
that corresponds  to $j_{c1}<j_{CF}<j_{c2}$. The quantities
$j_{c1}$ and $j_{c2}$ are the function of $\tilde{d}$ and
$\tilde{\nu}$. The dependences $j_{c1}$ and $j_{c2}$ on the
filling factor imbalance are shown in Fig. \ref{fig2} (these
dependences are symmetric with respect to $\tilde{\nu}$ and we
present them only for  $\tilde{\nu}>0$)  On can see from Fig.
\ref{fig2} that in balanced bilayers ($\tilde{\nu}=0$) the
diapason of currents at which stationary waves can be excited
shrinks to zero.

The existence of two critical currents $j_{c1}$ and $j_{c2}$
reflects the general situation in superfluid systems.  The
existence of Landau critical velocity (that in our case
corresponds to the current $j_{c1}$) is a common feature of
superfluid systems. While a single component uniform Bose-Einstein
condensate does not demonstrate the dynamical instability, more
complex superflud systems do. The examples are superfluid
Bose-Einstein condensates in a lattice \cite{lat1, lat2},
phase-separated two-component superfluid, where the components
have a common interface \cite{vv,ts1}, and two-component
superfluid mixtures (with penetrating components)
\cite{sf2,sf3,sf4}.

In general case  the Landau critical velocities are smaller than
ones that cause the dynamical instability, but in some cases the
system switches directly to the dynamical instability regime. As
was shown in \cite{sf2}, such a situation takes place in a
symmetric two-component Bose-Einstein condensate (two components
with the same characteristics) if the components flow with equal
velocities in opposite directions. In the latter case, in
similarity with a situation in balanced quantum Hall bilayers,
there is no range of velocities where stationary wave pattern may
emerge. Such a similarity can be understood as follows. In a
quantum Hall bilayer with $\nu_T=1$ the exciton transport can be
interpreted as counter-propagation of two exciton species of
different polarization, and in balanced bilayers such species move
in opposite directions with the same velocities.

The dynamical instability means that the magnetoexciton state is
unstable irrespectively to the interaction of magnetoexcitons with
an environment. The interaction with an environment may result in
a Landau (thermodynamic) instability in superfluids that develops
under the flow with the velocities larger than the Landau critical
velocities \cite{vv, ts2}. Phenomenologically, such type of
instability can be described by a modified Gross-Pitaevskii
equation that includes a dissipative term \cite{ts2}.
Microscopically, the dissipative term can be connected with an
interaction between the condensate and normal components in a
state where these two component are out of equilibrium with each
other, and the normal component interacts with an environment
\cite{zar, zar1}. As was shown in \cite{ts1,ts2}, the Landau
instability may reveal itself in a vortex nucleation at the
interface between two superfluid. In a typical situation the
Landau instability is much weaker than the dynamical instability.
In our study we do not take into account the Landau instability
effects implying that the system is in a regime where such effects
are small.

The quantity $K_{z\varphi}({\bf q})$ in the spectrum (\ref{e1})
contains the factor $\cos \theta_0$ which sign coincides with the
sign of the filling factor imbalance. Therefore at $\nu_1>\nu_2$
the stationary wave equation (\ref{st1}) can be satisfied for the
wave vectors that have negative projection on the direction of
${\bf Q}$ ($q_{x}<0$), and at $\nu_1<\nu_2$, for the wave vectors
with $q_{x}>0$. In systems that differ only by a sign of the
imbalance the stationary wave patterns will be the mirror images
of each other.

To be more specific we consider below  the case $\nu_1>\nu_2$. It
is convenient to parameterize the stationary wave crests by the
angle $\xi$ that determines the direction of the wave vector
counted from the $-x$ axis: ${\bf q}_{st} =(-q\cos\xi,q\sin\xi)$.
At $\nu_1>\nu_2$ the Mach cone is situated in the $x>0$ half-plane
and the Mach angle is $\alpha_M=\pi/2-\xi_M$, where $\xi_M$ is the
maximum angle of deviation of ${\bf q}_{st}$ from the $-x$
direction.

 Eq. (\ref{st1}) can be considered as
an implicit definition of the function $q_{st}(\xi)$. In general
case this equation has several solutions $q_{st,\lambda}(\xi)$,
where the index $\lambda$ numerates different families of
stationary waves.

To each wave vector ${\bf q}_{st}$  one can put in correspondence
the group velocity vector
\begin{equation}\label{vg}
    {\bf v}_{g}=\frac{1}{\hbar}\frac{\partial \Omega({\bf q})}{\partial
    {\bf q}}\Bigg|_{{\bf q}={\bf q}_{st}}.
\end{equation}
The polar angle $\chi$ that determines the direction of the group
velocity ${\bf v}_{g}$ relative to the $x$ axis is also the
function of $\xi$. The dependence $\chi(\xi)$ can be obtained from
Eq. (\ref{vg}) under substitution ${\bf
q}_{st}=(-q_{st}(\xi)\cos\xi,q_{st}(\xi)\sin\xi)$.

A point defect located at the origin emits stationary waves. Their
phase at the space point ${\bf r}$ is equal to
\begin{equation}\label{sp}
    \theta={\bf q}_{st} \cdot{\bf r}=-q_{st}(\xi) r \cos
    [\xi+\chi(\xi)].
\end{equation}
 The latter equation
gives the dependence $r(\xi,\theta)$. The coordinates of the wave
crests can be presented in the parametric form
\begin{eqnarray}\label{wcr}
    x(\xi,\theta)=-\frac{\theta \cos[\chi(\xi)]}{q_{st}(\xi)\cos
    [\xi+\chi(\xi)]},\cr
 y(\xi,\theta)=-\frac{\theta \sin[\chi(\xi)]}{q_{st}(\xi)\cos
    [\xi+\chi(\xi)]}.
\end{eqnarray}
The parameter $\xi$ belongs to the interval $|\xi|<\xi_M$. The
phase takes the values $\theta= \pm 2 \pi N+\theta_{in}$, where
$N=1,2,\ldots$, and $\theta_{in}$ is some initial phase that
cannot be determined in the kinematic approach  (for the
calculations we put $\theta_{in}=0$). The sign of $\theta$ is the
same for a given family of waves and can be found from the
condition $r(\xi,\theta)>0$ that is equivalent to $\theta \cos
[\xi+\chi(\xi)]<0$.

\begin{figure}
\begin{center}
\includegraphics{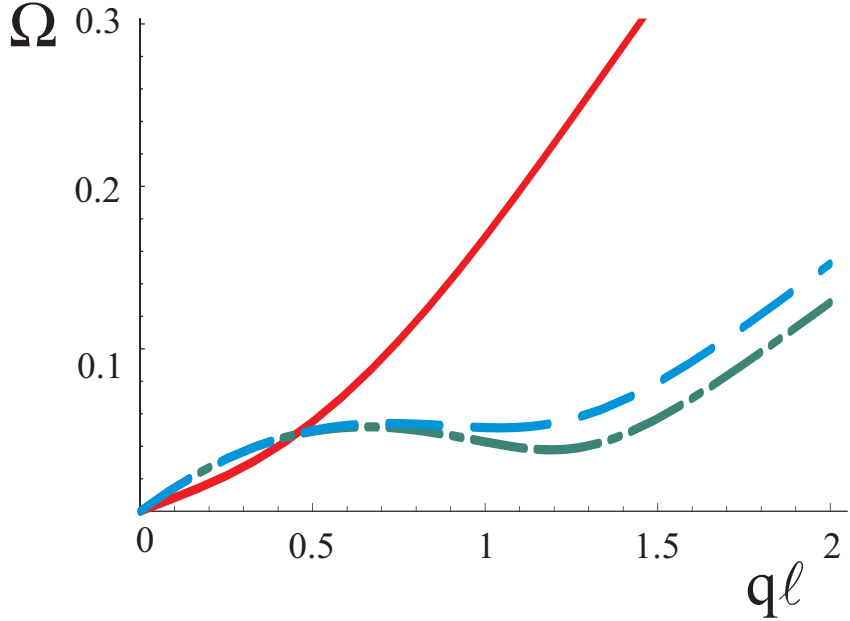}\\
  \caption{The spectrum of collective excitations (in
  $e^2/\varepsilon\ell$ units)
  at $d/\ell=0.3$ (solid line), $d/\ell=1.1$ (dashed line) and $d/\ell=1.2$
  (dash-and-dot line)
for the filling factor imbalance $\tilde{\nu}=0.25$.
 }\label{fig3}
\end{center}
\end{figure}

\begin{figure}
  \begin{center}
\includegraphics[width=7cm]{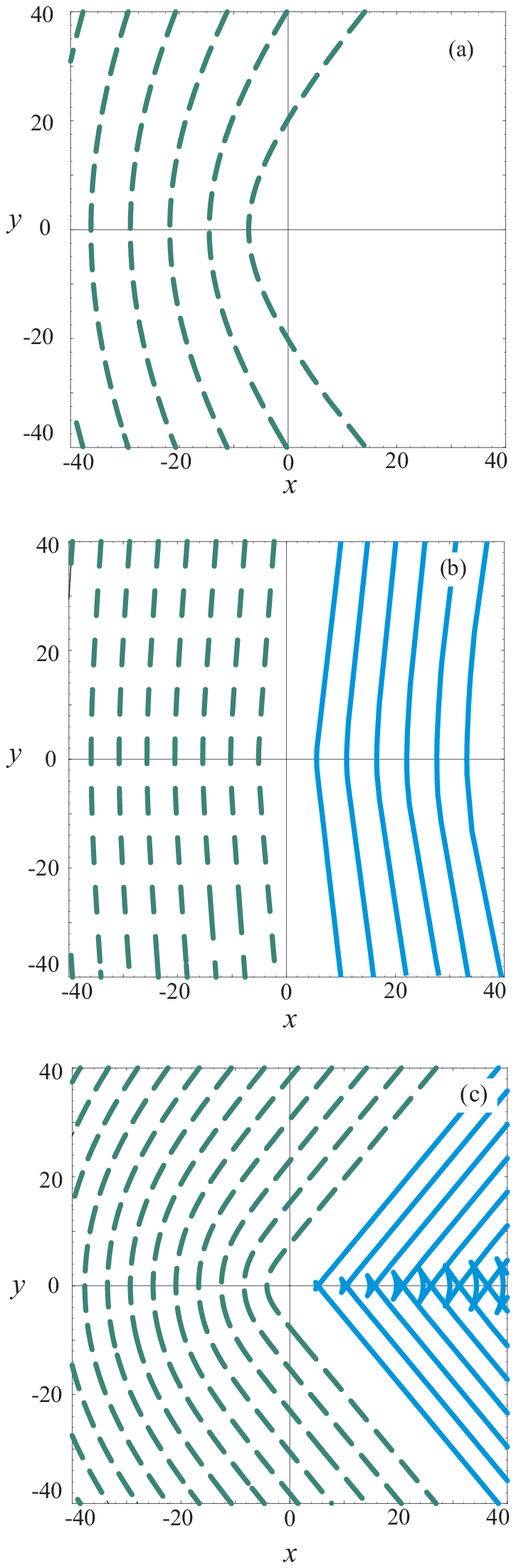}\\
  \caption{Stationary wave crests at $\tilde{\nu}=0.25$  and (a) --- $d/\ell=0.3$,
    $j_{CF}=4.5\cdot 10^{-3} e^3/\hbar\varepsilon \ell^2$; (b) --- $d/\ell=1.1$,
    $j_{CF}=2.2 \cdot 10^{-3}  e^3/\hbar\varepsilon \ell^2$; (c) ---  $d/\ell=1.2$,
    $j_{CF}=1.9 \cdot 10^{-3} e^3/\hbar\varepsilon\ell ^2$.
    The family located outside (inside) the Mach cone is shown
     by dashed (solid) lines.
    The spatial scale is $1=\ell$.}\label{fig4}
  \end{center}
\end{figure}

As was mentioned in Introduction, the location of stationary waves
is determined by the type of dispersion of the collective mode. In
the case considered the dispersion depends on the parameter
$\tilde{d}$. Typical dispersion curves at $Q=0$ are shown in Fig.
\ref{fig3}. One can see that at rather small $\tilde{d}$ the
dispersion is positive at all $q$. At larger $\tilde{d}$ the
dispersion becomes negative at intermediate $q$. At $\tilde{d}$
close to critical one the spectrum contain a minimum at finite
$q$. The wave crests obtained for these three cases are shown in
Fig. \ref{fig4}. One can see that in the case of small $\tilde{d}$
the stationary wave pattern is similar to one that was predicted
and observed in weakly non-ideal superfluid Bose gases, while at
larger $\tilde{d}$ an additional family of stationary waves
emerges inside the Mach cone. The minimum in the spectrum reveals
itself in an appearance of cusps on the crest lines.

\section{Stationary wave patters in a superfluid magnetoexciton gas}
\label{s3}

To compute the amplitudes of stationary waves we will use the
dynamical approach \cite{12}. Stationary waves emerge as a
response on an obstacle that was appeared in the past. The
casuality principle can be realized by the following choice for
the Hamiltonian of interaction with the obstacle
\begin{equation}\label{hint}
    H_{int}=\int d^2 r e^{\eta t} \sum_i V_i ({\bf r})   \rho_i({\bf r}),
\end{equation}
where $V_i({\bf r})$ ($i=1,2$) is the interaction potential,
$\rho_i({\bf r})$ is the electron density operator, and
 $\eta=+0$. We consider the point obstacle $V_i({\bf r})=U_i \delta({\bf
 r})$ and imply that the interaction ({\ref{hint}) does not
 change the sum of local filling factors of the layers.

In the state (\ref{dcs}) the energy of the interaction
(\ref{hint}) reads as
\begin{equation}\label{hint1}
    E_{int}=\frac{U_z e^{\eta t}}{S}\sum_{q}\sum_X \cos \theta_X \frac{1+e^{i q Q_y
    \ell^2}}{4}e^{-i q X-\frac{q^2 \ell^2}{4}},
\end{equation}
where $U_z=U_1-U_2$.  The fluctuating part of (\ref{hint1})
expressed in terms of $m_z(q)$ has the form
\begin{equation}\label{hint2}
 E_{int}^{fl}=\frac{U_z e^{\eta t}}{\sqrt{2\pi \ell^2 S}}\sum_{q} m_z(q) \frac{1+e^{i q Q_y
 \ell^2}}{2}e^{-\frac{q^2 \ell^2}{4}}.
\end{equation}

The interaction (\ref{hint2}) modifies the equations of motion:
\begin{eqnarray} \label{13-a}
\hbar\frac{d {\varphi}(q)}{d t}=2 K_{zz} (q) {m}_z(q)- i
K_{z\varphi}(q) {\varphi}(q)+\frac{U_z e^{\eta t}}{\sqrt{2\pi
\ell^2 S}}\frac{1+e^{- i q Q_y
 \ell^2}}{2}e^{-\frac{q^2 \ell^2}{4}}
  ,\cr
\hbar\frac{d {m}_z(q)}{d t}=-\frac{1}{2} K_{\varphi\varphi}(q)
{\varphi}(q)  -  i K_{z\varphi}(q) {m}_z(q).
\end{eqnarray}

Taking the partial solution of (\ref{13-a}) at $t=0$,  we obtain
the following expression for $m_z(q)$
\begin{equation}\label{mzp}
    m_z(q)=-\frac{U_z}{4\sqrt{2\pi
\ell^2 S}}\frac{K_{\varphi\varphi}(q)(1+e^{-i q
Q_y\ell^2})e^{-\frac{q^2
\ell^2}{4}}}{K_{zz}(q)K_{\varphi\varphi}(q)-(K_{z\varphi}(q)-i\hbar\eta)^2
}.
\end{equation}
Using the  inverse Fourier transformation for $m_z(q)$
\begin{equation}\label{it}
    \cos \theta_X=\cos \theta_0+2\sqrt{\frac{2\pi\ell^2}{S}}\sum_q
     m_z(q)e^{i q X},
\end{equation}
the expression for the Fourier component of the electron density
difference
\begin{equation}\label{it1}
    \rho_1({\bf q})-\rho_2({\bf q})=\delta_{q_y,0}\sum_X \cos
    \theta_X \frac{1+e^{i q_x Q_y
 \ell^2}}{2} e^{-i q_x X -\frac{q^2\ell^2}{4}},
\end{equation}
and taking the inverse Fourier transformation of (\ref{it1})
\begin{equation}\label{it3}
 \Delta\rho({\bf r})=\rho_1({\bf r})-\rho_2({\bf r})=\frac{1}{S}\sum_{\bf q}
 \left[\rho_1({\bf q})-\rho_2({\bf q})\right] e^{i{\bf q}\cdot{\bf
 r}},
\end{equation}
we obtain
\begin{equation}\label{it4}
\Delta\rho({\bf r})=\frac{\cos \theta_0}{2\pi\ell^2}-\frac{
U_z}{2\pi\ell^2}\frac{1}{2S}\sum_{\bf q} e^{i{\bf q}\cdot{\bf
r}}\delta_{q_y,0}\frac{K_{\varphi\varphi}(q)[1+\cos(q_x
Q_y\ell^2)]e^{-\frac{q^2
\ell^2}{2}}}{K_{zz}(q)K_{\varphi\varphi}(q)-(K_{z\varphi}(q)-i\hbar\eta)^2}
\end{equation}
The first term in Eq. (\ref{it4}) is the uniform electron density
imbalance $\Delta\rho_0$.  The nonuniform part
$\tilde{\rho}=\Delta\rho-\Delta\rho_0$ describes the density
fluctuations caused by the stationary waves. Eq. (\ref{it4})
yields the contribution of the modes with the wave vectors
directed along the $x$ axis. The contribution of modes with
general ${\bf q}$ can be obtained by rotation of the coordinate
axes. Summing the contribution of all ${\bf q}$ we find
\begin{equation}\label{it5}
\tilde\rho({\bf r})=-\frac{ U_z}{2\pi\ell^2}\frac{1}{2S}\sum_{\bf
q}e^{i{\bf q}\cdot{\bf r}} \frac{K_{\varphi\varphi}({\bf
q})[1+\cos(|{\bf q}\times {\bf Q}|\ell^2)]e^{-\frac{q^2
\ell^2}{2}}}{K_{zz}({\bf q})K_{\varphi\varphi}({\bf
q})-(K_{z\varphi}({\bf q})-i\hbar\eta)^2}.
\end{equation}
 Replacing the sum
with the integral and using the symmetry properties of (\ref{it5})
we obtain
\begin{eqnarray}\label{mr3}
 \tilde{\rho}(\mathbf{r})=-\frac{ U_z}{2\pi l^2}
    \frac{1}{(2 \pi)^2}\textrm {Re}
    \int_{\frac{-\pi}{2}}^{\frac{\pi}{2}}d\xi \int_0^\infty d q
    e^{-i q r \cos(\xi+\chi)}
I(q,\xi),
    \end{eqnarray}
where
\begin{equation}\label{i7}
    I(q,\xi)=\frac{q
K_{\varphi\varphi}(q,\xi)[1+\cos (q   Q
\ell^2\sin\xi)]e^{-\frac{q^2
l^2}{2}}}{K_{zz}(q,\xi)K_{\varphi\varphi}(q,\xi)-(K_{z\varphi}(q,\xi)-i\hbar\eta)^2},
\end{equation}
 and $\chi$ is the polar
angle for the vector ${\bf r}$.

The integral over $q$  in Eq. (\ref{mr3}) is evaluated from the
residue theorem. The integration contour is chosen as shown in
Fig. \ref{fig5}. The contour 1 in Fig. \ref{fig5} corresponds to
the case where the observation point is outside the Mach cone, and
the contour 2, inside the Mach cone. The integral along the
imaginary axis  yields the contribution of order $(r \ell)^{-2}$,
while the integral along the real axis - the contribution of order
$(r \ell)^{-1/2}$. We specify the case $r \ell\gg 1$ for which the
first contribution can be neglected. In this approximation
\begin{eqnarray}\label{mr3-1}
 \tilde{\rho}(\mathbf{r})\approx \frac{ U_z}{2\pi l^2}
    \frac{1}{2\pi}\textrm {Im} \int_{\frac{-\pi}{2}}^{\frac{\pi}{2}}d\xi
\sum_\lambda  s_\lambda e^{-i q_\lambda(\xi) r \cos(\xi+\chi)}
{\rm Res}_{q=q_\lambda(\xi)}[I(q,\xi)],
\end{eqnarray}
where $q_\lambda(\xi)$ are the poles of $I(q,\xi)$, and the factor
\begin{equation}\label{tet}
s_\lambda=\left\{%
\begin{array}{ll}
    1, & \hbox{$\cos(\xi+\chi)  {\rm Im}(q_\lambda(\xi))
    <0$} \\
    0, & \hbox{$\cos(\xi+\chi)  {\rm Im}(q_\lambda(\xi))
    >0$} \\
\end{array}%
\right.
\end{equation}
indicates whether is the pole $q_\lambda(\xi)$ inside or outside
the contour of integration Fig. \ref{fig5}.

\begin{figure}
  \begin{center}
\includegraphics{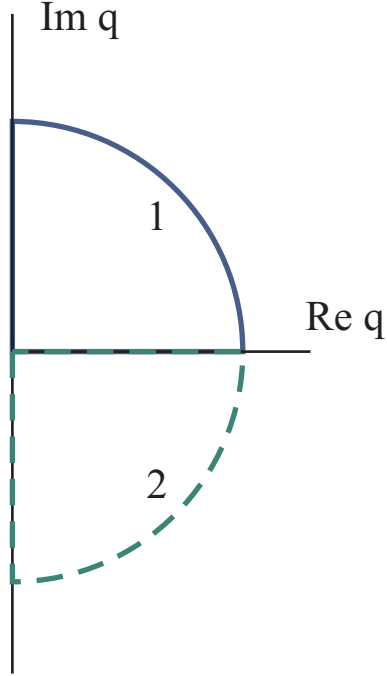}\\
  \caption{The integration contours
  used for the computation of the integral (\ref{mr3})}\label{fig5}
  \end{center}
\end{figure}

The integral over $\xi$ is evaluated in the stationary phase
approximation
\begin{eqnarray}\label{mr5}
 \tilde{\rho}(\mathbf{r})\approx \frac{ U_z}{2\pi l^2}
    \sum_\lambda \sum_{\xi_i}
    A_\lambda(\xi_i)\frac{\sin\left[q_\lambda(\xi_i)r \cos(\chi+\xi_i)
    +\frac{\pi}{4}{\rm sgn}
    [g_\lambda(\xi_i)]\right]}{\sqrt{r}}.
\end{eqnarray}
Here $\xi_i$ are the solutions of the stationary phase equation
\begin{equation}\label{stph}
    \frac{d}{d \xi}\left[q_\lambda(\xi)\cos(\chi+\xi)\right]=0,
\end{equation}
the function $g_\lambda(\xi)$ is defined as
\begin{equation}\label{gk}
g_\lambda(\xi)= \frac{d^2}{d
\xi^2}\left[q_\lambda(\xi)\cos(\chi+\xi)\right],
\end{equation}
and the amplitude $A_\lambda(\xi)$ is equal to
\begin{equation}\label{amp}
   A_\lambda(\xi)=-s_\lambda\sqrt{\frac{1}{2\pi
   |g_\lambda(\xi)|}}{\rm Res}_{q=q_\lambda(\xi)}[I(q,\xi)].
\end{equation}

One can check that Eq. (\ref{mr5}) up to the initial phase yields
the same wave crests as determined by the parametric equations
(\ref{wcr}). Indeed, in the limit $\eta\to 0$ the poles
$q_\lambda(\xi)$ coincide with the stationary wave vectors
determined by Eq. (\ref{st1}). The stationary phase equation
(\ref{stph}) determines the multi-valued function $\xi_i(\chi)$
that is reciprocal to the function $\chi(\xi)$ defined by Eq.
(\ref{vg}). Let us prove the last statement. Considering $q$ and
$\xi$ as independent variables one finds
\begin{eqnarray}\label{pr1}
    \frac{\partial\Omega}{\partial q_x}=-\frac{\partial\Omega}{\partial
    q}\cos\xi-\frac{\partial\Omega}{\partial
    \xi}\frac{\sin \xi}{q}\cr
\frac{\partial\Omega}{\partial q_y}=\frac{\partial\Omega}{\partial
    q}\sin\xi+\frac{\partial\Omega}{\partial
    \xi}\frac{\cos \xi}{q}
    \end{eqnarray}
Taking into account the relation
$$\frac{\partial\Omega}{\partial q}\Bigg|_{q=q_{st}(\xi)}
q_{st}'(\xi)+\frac{\partial\Omega}{\partial
\xi}\Bigg|_{q=q_{st}(\xi)}=0$$ and Eqs. (\ref{vg}),(\ref{pr1}) we
obtain
\begin{equation}\label{pr2}
    \tan \chi =\frac{\sin \xi-\frac{q'}{q}\cos\xi}{-\cos\xi+\frac{q'}{q}\sin\xi}
\end{equation}
that yields
\begin{equation}\label{pr3}
\tan (\chi+\xi)=\frac{q_{st}'(\xi)}{q_{st}(\xi)}
\end{equation}
Eq. (\ref{pr3}) is equivalent to Eq. (\ref{stph}). Note that Eq.
(\ref{pr3}) gives the explicit expression for the function
$\xi(\chi)$ that can be used instead of the implicit expression
(\ref{vg}) for the computation of the wave crest lines.

The stationary wave density pattern given by Eq. (\ref{mr5}) is
shown in Fig. \ref{fig6}. For the computations we use the same
parameters as in Fig. \ref{fig4}c.

\begin{figure}
\begin{center}
\includegraphics{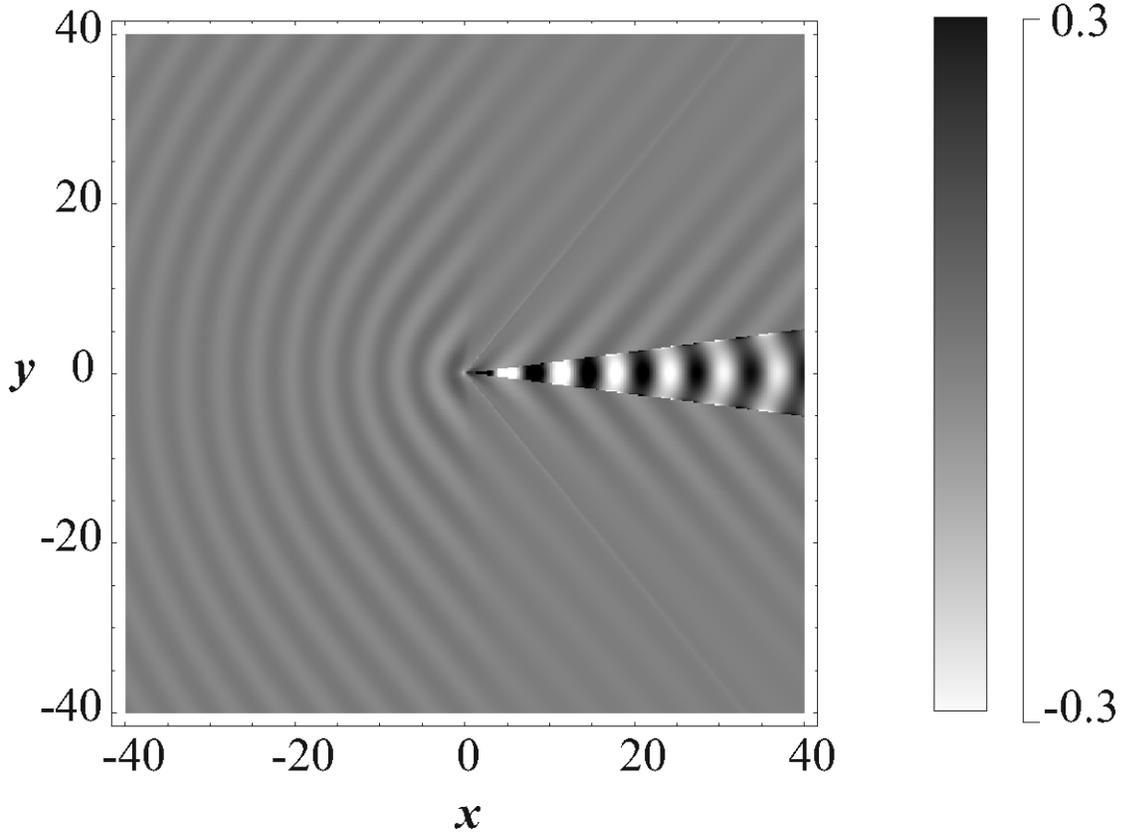}\\
  \caption{The stationary wave density pattern
   for the spectrum with a minimum at finite $q$. The density difference is given in $U_z/2\pi \ell^2
E_C$ units, where $E_C=e^2/\varepsilon \ell$ is the Coulomb
energy. The spatial scale is $1=\ell$.}\label{fig6}
\end{center}
\end{figure}

\section{Stationary density waves in a graphene-dielectric-graphene sandwich}
\label{s4}

Stationary waves in quantum Hall bilayers reveal itself in
spatially non-uniform local electrical fields that emerge due to
variations of electron densities in the layers. The period of
stationary waves is proportional to the magnetic length. For
typical magnetic fields used for the study of magnetoexciton
superfluidity in GaAs heterostructures magnetic length is of order
of 10 nm that corresponds to the stationary wave period less than
$100$ nm (see Fig. \ref{fig6}). For the observation of stationary
waves an
 electrostatic field detector should be located  close to the
electron layer and its spatial resolution should be smaller than
the stationary wave period.

In this section we discuss advantages of use bilayer graphene
structures for the observation of stationary waves in the
magnetoexciton gas. The problem of exciton superfluidity in
double-layer graphene structures was studied theoretically in a
number of papers \cite{19,22,23,24,24a,25}.  Here we follow the
paper \cite{19}, where the theory of magnetoexciton superfluidity
in graphene bilayers was developed.

In difference with quantum wells, the graphene layers can be
located at the surface of the system. It was reported recently on
the creation of a bilayer graphene structure that has one open
layer and separate contacts in each layers \cite{21}.

Another advantage of graphene systems is much larger energy
distance between Landau levels in comparison with one for quantum
wells in GaAs heterostructures. That allows to use  smaller
magnetic fields. Smaller field correspond to larger magnetic
length and larger spatial period of stationary waves in absolute
units.

Let us discuss this question in a more quantitative way. The
energies of Landau levels in graphene are given by the expression
\cite{rmp}
\begin{equation}\label{llg}
    E_{\rm LL}(N)={\rm sgn}(N)\frac{\hbar v_F\sqrt{2|N|}}{\ell},
\end{equation}
where $N$ is integer number, and $v_F=10^6$ m/s is the Fermi
velocity for graphene. Landau levels in graphene have the
additional fourfold degeneracy (four components correspond to
different combinations of spin and valley indices). In a
free-standing graphene the $N=0$ level is half-filled, but in
difference with the single component systems magnetoexciton
superfluidity in the graphene bilayers is not possible at
half-filling \cite{19}.

The situation is similar to ones for $\nu_T=2$ double quantum
wells \cite{26}. Electrostatic field applied perpendicular to the
graphene layers may create an imbalance of filling of $N=0$ Landau
level in the layers 1 and 2. We consider the cases
$\nu_1=2+\nu_{a}$, $\nu_2=2-\nu_{a}$, and $\nu_1=3+\nu_{a}$,
$\nu_2=1-\nu_{a}$ ($0<\nu_a<1$). Here the filling factors are
defined as the average number of filled quantum states in a given
level per the area $2\pi\ell^2$. The quantities $\nu_{a1}=\nu_a$
and $\nu_{a2}=1-\nu_a$ give the filling factors  for the so called
active component: only for one (active) component the modulus of
the order parameter for the electron-hole pairing is nonzero:
$|\Delta_a|=\sqrt{\nu_a(1-\nu_a)}$.

The eigenfunction for the zero Landau level in graphene coincides
with the eigenfunction for free electron gas. That is why the
dynamics of the active component in the bilayer graphene system is
described by basically the same equations that were used in the
previous section.

Here we consider the  graphene-dielectric-graphene sandwich
structure with two open graphene layers (two layers separated by a
dielectric layer which thickness is much larger than the distance
between graphenes in graphite). For such a structure the Fourier
components of the intralayer and interlayer Coulomb interaction
read as
\begin{eqnarray}\label{v11}
    V_S(q)=\frac{4 \pi e^2}{q}
    \frac{(\varepsilon+1)e^{ q d}+(\varepsilon-1)e^{- q d}}
    {(\varepsilon+1)^2e^{ q d}-(\varepsilon-1)^2e^{- q d}},\cr
V_D(q)=\frac{2 \pi e^2}{q}\frac{4
\varepsilon}{(\varepsilon+1)^2e^{ q d}-(\varepsilon-1)^2e^{- q
d}}.
\end{eqnarray}
The dielectric constant for the environment $\varepsilon_{en}=1$
is assumed.

The empty states in a partially filled Landau level can be
considered as holes if the Coulomb interaction energy is smaller
than the energy distance between Landau levels.
 Let us use the inequality
\begin{equation}\label{ine}
   F_S(0)< E_{LL}(1)-E_{LL}(0)=\frac{\sqrt{2}\hbar v_F}{\ell}
\end{equation} as a quantitative condition of smallness of the Coulomb
interaction. The quantity $F_S(0)$ determines the strength of the
intralayer exchange interaction. The right hand part of Eq.
(\ref{ine}) is the distance between the active ($N=0$) and the
nearest passive ($N=\pm 1$) Landau level for graphene. For the
geometry considered in \cite{19} (two graphene layers embedded in
a dielectric matrix) the quantity
$F_{S}(0)=\sqrt{\pi/2}e^2/\varepsilon\ell$ and the inequality
(\ref{ine}) yields the restriction only of the dielectric constant
($\varepsilon>2$).
 For the graphene-dielectric-graphene
sandwich the intralayer exchange energy depends on $\tilde{d}$:
\begin{equation}\label{fs0}
    F_S(0)=\frac{e^2}{\ell}\int_0^\infty d x
    e^{-\frac{x^2}{2}}
    \frac{2[(\varepsilon+1)e^{x\tilde{d}}+(\varepsilon-1)e^{-x\tilde{d}}]}
    {(\varepsilon+1)^2e^{x\tilde{d}}-(\varepsilon-1)^2e^{-x\tilde{d}} }
\end{equation}
and the inequality (\ref{ine}) yields the restriction
$\tilde{d}>\tilde{d}_{c1}(\varepsilon)$, where the function
$\tilde{d}_{c1}(\varepsilon)$ is given implicitly by the equation
$F_S(0)={\sqrt{2}\hbar v_F}/\ell$. On the other hand, the
condition for the excitation spectrum be real valued yields
another restriction on the parameter $\tilde{d}$:
$\tilde{d}<\tilde{d}_{c}(\tilde{\nu}_a, \varepsilon)$, where
$\tilde{\nu}_a=\nu_a-1/2$ is the imbalance for the active
component. In difference with the situation shown in Fig.
\ref{fig1} the critical distance $\tilde{d}_{c}$ for the
graphene-dielectric-graphene sandwich structure
 depends on the dielectric constant. The
dependences $\tilde{d}_{c}(\varepsilon)$ and
$\tilde{d}_{c1}(\varepsilon)$  are shown in Fig. \ref{fig7}. One
can see that $\tilde{d}_{c1}<\tilde{d}_{c}$  for
$\varepsilon\gtrsim 4$ and under appropriate choice of $d/\ell$
one can satisfy the both restriction. At fixed $d$ the magnetic
length is restricted from above and from below, but the diapason
of allowed $\ell$ can be shifted by a change of $d$. Thus, one can
adjust the parameters to make the period of stationary waves
appropriate for the observation.

\begin{figure}
  \begin{center}
\includegraphics{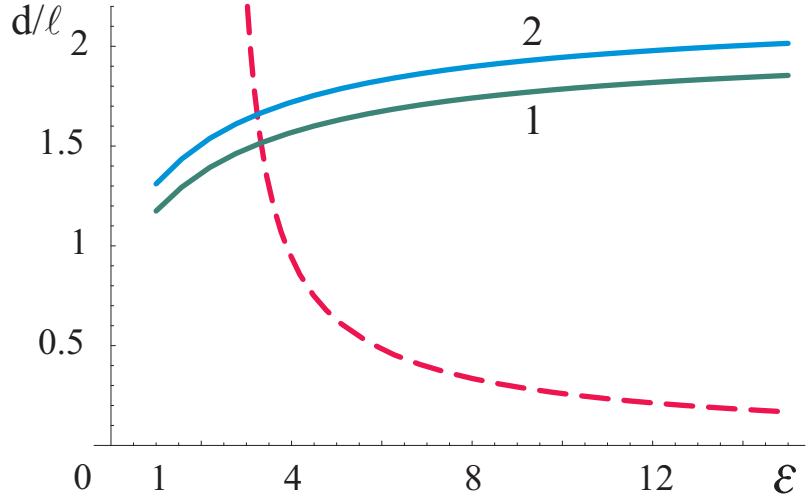}\\
  \caption{Critical interlayer distance $d_c$ (solid lines 1 -- $\tilde{\nu}_a=0$,
  solid lines 2 -- $\tilde{\nu}_a=0.25$) and
  $d_{c1}$ (dashed line) for the graphene-dielectric-graphene
  sandwich structure.}\label{fig7}
  \end{center}
\end{figure}

In Fig. \ref{fig8} we present the stationary wave density pattern
computed for a graphene-dielectric-graphene structure.  The cases
of one family and two families without cusps are shown. One can
see from the Figs. \ref{fig6} and \ref{fig8}a,b that the amplitude
of stationary waves grows up when $d$ approaches to $d_c$.

\begin{figure}
\begin{center}
\includegraphics[width=16 cm]{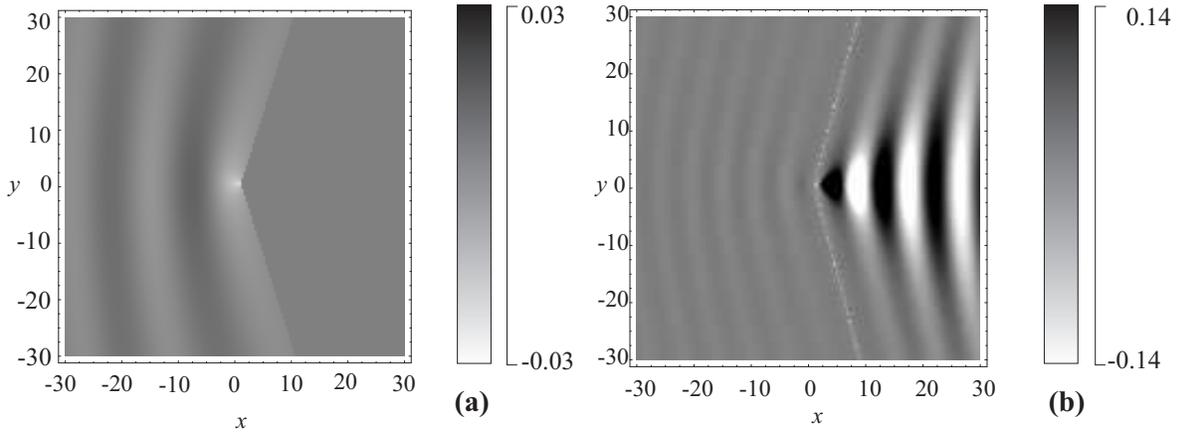}\\
  \caption{Stationary wave patterns in a
  graphene-dielectric-graphene sandwich
  structure with $\varepsilon=15$
  and $\tilde{\nu}_a=0.25$.  (a) -- $d/\ell=1.1$,
   $j_{CF}=1.1\cdot 10^{-2} e^3/\hbar\varepsilon \ell^2$;
   (b) -- $d/\ell=1.5$, $j_{CF}= 6.9 \cdot 10^{-3} e^3/\hbar\varepsilon
   \ell^2$}\label{fig8}
\end{center}
\end{figure}

\section{Conclusion}

In conclusion, we wave studied stationary waves in a
magnetoexciton gas in quantum Hall bilayers. Stationary waves are
static excitations of the difference of electron densities in the
layers. They are induced by the counter-propagating electrical
currents flowing in a system with an obstacle. It is found that
such waves can be excited only in imbalanced bilayers in a certain
range of currents. The stationary wave pattern is modified
qualitatively under a variation of the ratio of the interlayer
distance to the magnetic length $d/\ell$. At small $d/\ell$ the
pattern is similar to one that emerges in a superfluid weakly
non-ideal Bose gas: stationary waves are located  outside the Mach
cone. At larger $d/\ell$ an additional family of waves is located
inside the Mach cone appears. Under further increase of $d/\ell$
cusps are developed in the crests that correspond to the second
family. The amplitudes of stationary waves grow up under increase
of $d/\ell$. A typical spatial period  of stationary waves is
several magnetic lengths. A convenient system for the observation
of stationary waves is a graphene-dielectric-graphene structure.
In such a system the spatial period of stationary waves can be
rather large and one can put a detector close to the graphene
layer.

This study is supported by the Ukraine State Program
"Nanotechnologies and nanomaterials" Project No 1.1.5.21.

\end{document}